# Thinking Out Loud and e-Health for Coordinated Care – Lessons from User Requirements Gathering in the 4C Project

**Leonie Ellis**
School of Engineering and ICT
University of Tasmania
Hobart, Australia
Email: Leonie.Ellis@utas.edu.au

**Colleen Cheek**
Rural Clinical School
University of Tasmania
Hobart, Australia
Email: Colleen.Cheek@utas.edu.au

**Paul Turner**
School of Engineering and ICT
University of Tasmania
Hobart, Australia
Email: Paul.Turner@utas.edu.au

## Abstract

e-Health is a core part of Australia's strategy to address rising costs and changing demands for healthcare services. That over $1bn has been spent and only 6% of Australians registered to the personally controlled electronic health record suggests user challenges remain. Evidence confirms the benefits from involving users in systems development there is a need for examples of how to engage effectively in healthcare settings. This research describes the use of an agile development methodology combined with the 'Thinking Out Loud' technique to deliver a solution that exceeded user requirements in supporting a new model of care. The 4C project solution proposed to connect Aged Care institutions with general practices, hospitals and specialist services in Tasmania's north-west region. It was underpinned by a design incorporating three spheres of participation. As a trial project for the PCEHR it remains unclear why lessons learned appear not to have been deployed more explicitly in the national roll-out.

## 1  INTRODUCTION

Governments around the world continue to struggle with burgeoning health budgets to respond to changing demands resulting from increasing chronic disease prevalence, an ageing population, advancing technology, and increasing consumer expectations for care (Commonwealth of Australia 2010).  The benefits of sharing electronic health information have been discussed over many years, with private health providers and government programs generously investing in the infrastructure to make this possible. In Australia the National e-Health Strategy outlined the need for a coordinated and aligned approach to developing e-health solutions to ensure integration of efforts and expenditure scaled across the continuum of care.

*"E-Health will enable a safer, higher quality, more equitable and sustainable health system for all Australians by transforming the way information is used to plan, manage, and deliver health care services."* (Commonwealth of Australia 2010)

In the UK, the National Health Service recognized the need in 2006 to embrace a more patient-led approach to delivering services as a means of directing more efficient and effective use of health budgets (Cayton 2006).  The challenge for health information systems is to move beyond information access to more informed involvement, not just accessibility of health information, but transforming models of care with greater consumer participation.

In 2009, to help promote discussion of these patient-centred research agendas, the United States National Science Foundation sponsored workshops among leading academics, researchers, government staffers, and graduate students (Shneiderman 2011). One of the identified themes for the discussion panels was health and wellness possibilities. The panel utilised a framework of three spheres of participation identified by the National Health Service as essential for delivering healthcare in the 21st century. These comprise:

- a **population sphere** of public health officials and communities seeking best policy to protect all citizens equitably;

- **a personal sphere** where citizens have tools that empower them to pursue the best health strategies for themselves and their families; and

- a **clinical sphere** in which practitioners can communicate with each other on individual cases and best practice.

## 2 AUSTRALIAN PERSONALLY CONTROLLED EHEALTH RECORDS (PCEHR) AND THE 4C PROJECT

The Personally Controlled eHealth Record (PCEHR) system is arguably the largest green-field health infrastructure project in Australia since the introduction of Medicare. New Commonwealth legislation on PCEHRs and Healthcare Identifiers has been passed, and registrations commenced in July 2012. The PCEHR System will provide the infrastructure required for every Australian to have an electronic record containing the key components of their medical history. Once created, an individual's PCEHR will be stored in a network of connected systems, with the information being available to the individual and their authorised healthcare providers. The network will facilitate sharing, at a national level, of important information about an individual's healthcare with that individual's authorised healthcare providers (NeHTA 2011). This new system went live on 1st July 2012, but to date only 6% of Australian citizens have registered to use the system despite a huge investment of commonwealth funds (Pearce and Bainbridge 2014).

Significantly, prior to the PCEHR commencement twelve federally funded PCEHR conformant implementation projects (Wave 2) were funded in 2011 to trial different aspects of the PCEHR system. The Cradle Coast Connected Care (4C) project was chosen by the Commonwealth Government as one of the Wave 2 eHealth sites. It was to be an early example of an electronic health record system and from the outset endeavoured to use a health information system for meaningful use.

The 4C electronic health record was developed to align with a new model of supportive and palliative care being championed by the local health service, Tasmania Health Organisation-North West.

### 2.1 The 4C EHR and Advance Care Planning for cross boundary care

Many people fear a loss of autonomy, dignity and the ability to make their preferences known after they have lost decision-making capacity. Advance care directives (ACDs) are a document and a process by which a person, often with the involvement of friends, family and their treating team, plans for the types of end of life (EoL) care they wish to have after the person loses the capacity to be involved in making those decisions for themself.

Paper-based ACDs have been used for some time. Unfortunately, health professionals providing EoL care to a patient, especially in acute settings, may not be aware that the patient has an ACD. Prior to the 4C project the ACDs were not registered nor easily made available to the health professional. In addition, cross-boundary care requires articulation of goals of care that can be readily communicated with general practitioners, the Primary Health Care team, the family, acute hospital services, and specialist palliative care services. The 4C system aimed to support cross-boundary care required for the 4C patient cohort by establishing a regional repository that would connect RACFs with general practice, acute hospital services, the after hours general practitioner telephone advice service, and allied health providers such as community pharmacy. The 4C system proposed to keep and maintain care plans, best practice guidelines for patient symptom management, individualised as well as generic guidelines for unexpected deterioration and crisis management, legal documents (such as a statement of wishes) and Enduring Guardian (person registered in Tasmanian law to make decisions on behalf of the incapacitated individual).

The three spheres of participation provides a framework to explore these supportive and palliative care issues in context:

### 2.2 The Population Sphere

Higher standards of health have resulted in more people living longer - the proportion of those living beyond 60 years has increased and this proportion is predicted to increase over the next 20 years. While medical advances have increasingly allowed life to be prolonged, this comes at an extra expense to society. A Quality of Life Index (Economist Intelligence Unit 2010) devised to rank countries according to their provision of end-of-life (EoL) care assessed the United Kingdom as leading the world in overall quality of EoL care. While Australia is ranked second in quality, we take top place with the costs associated with the provision of EoL care. In comparing the costs of EoL care, the United Kingdom ranks 18th (Table 1). The explanation offered for the higher costs in Australia is the stronger focus on hospital-based medical care.

| Country | Rank (Quality) | Rank (Cost) |
|---|---|---|
| Australia | 2 | 1 |
| United Kingdom | 1 | 18 |

*Table 1. Ranking end-of-life care (of 40 countries listed) – (from Economist Intelligence Unit 2010)*

From a comparison of major Organisation for Economic Co-operation and Development (OECD) countries, neighbours and trading partners, Australia has the 3rd highest life expectancy, at 81.5 years (ABS 2010). It is projected there will be 1.8 million Australians aged 85+ by 2050, which means the Aged Care residents in the 85+ age group will rise from the current 87,000 to over 400,000 by 2050 (Australian Productivity Commission 2011). Issues of concern include staff shortages, high staff turnover, lack of palliative care skills, and quality of care.

### 2.3 The Personal Sphere

Higher spending is not always associated with higher quality care, better access to care, or better health outcomes or satisfaction (Davies and Higginson 2004). Recent Australian debate about the right of citizens to choose when and how to die suggests individual preferences are at odds with traditional healthcare models available to them. A body of evidence is mounting to show that older people suffer unnecessarily because of widespread under-assessment and inappropriate over-treatment or under-treatment of their problems (Davies and Higginson 2004). They experience multiple problems and disabilities and require more complex care. One factor for this is that many suffer multiple comorbidities, multiple chronic illnesses, such as the frail aged, with heart failure and dementia (Davies and Higginson 2004). These people have not had ready access to a combined supportive and palliative approach to their care in the few last years of life (Senate Community Affairs Committee 2012).

The World Health Organisation suggests the inconsistency between preference and action is associated with higher health care costs and that an ageing population does not necessarily mean that the cost of care for people in the last years of their life should overwhelm health service funding.

*"It may therefore not be the role of health care to seek a cheap solution to the issues that technology and ageing present, but to provide packages of care for people in different situations that properly take account of their wishes"* (Davies and Higginson 2004, p12).

### 2.4 The Clinical Sphere

As populations age, the pattern of diseases that people suffer and die from also changes. Increasingly, more people die as a result of serious chronic diseases such as heart disease, and cerebrovascular disease including stroke, respiratory disease and cancer. Deaths due to Dementia and Alzheimer's disease have doubled in Australia in the last 10 years (ABS 2010). Cognitive and behavioural impairment arising from Dementia complicates assessment especially, for example, the timely recognition of pain.

Specialist Palliative Care Services have traditionally focused their limited resources toward the needs of people dying from the complexities of advanced cancer (Senate Community Affairs Committee 2012). Figure 1 shows while cancer is the cause of one third of all deaths in Australia, 67% of deaths are from causes other than cancer and sudden and accidental death. In Registered Aged Care Facilities (RACF) the incidence of cancer is likely to be much less than the 25% in the general population, with a greater proportion with chronic advanced disease.

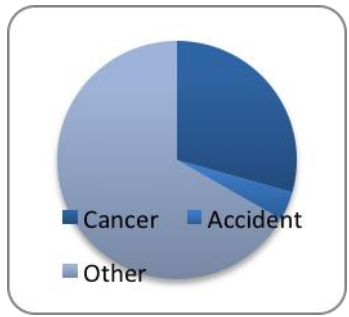

*Figure 1 - Causes of Death in Australia in 2008 (ABS 2008)*

Due to uncertainty of the timing of death, many medical and nursing professionals have been reluctant to engage patients in opportunities to be involved in decisions around their care, to address realistic goals of care and advance care planning (Engelhardt et al. 2009; Happ et al. 2002; Levin et al. 2010).

### 2.4.1 Objectives of 4CEHR within the 3 spheres of participation framework:

The objectives of the 4CEHR in the population sphere were:

- by providing timely shared electronic health information, better use limited acute hospital resources by reducing the use of medically futile interventions, reducing overly burdensome interventions, and offering care that is beneficial to the resident and is, appropriate to the stage of illness, prognosis, with achievable and realistic goals of care;
- to implement a regional PCEHR – compliant repository utilising national infrastructure and standards where available;
- to test and advise on components of the national PCEHR infrastructure; and
- to provide a tool that supports planning for expected and unexpected deteriorations and for EoL Life care to reduce crisis admissions, reduce avoidable hospitalisations, and reduce length of stay for appropriate admissions, enabling more to die in the RACF.

The objectives of the 4CEHR in the personal sphere were:

- to provide information via an electronic portal to improve involvement of residents and families in decision-making about their care, identifying preferences, wishes, and performing ACP so that the care received is aligned with the resident's wishes;
- through provision of a 4CEHR, to improve the timely delivery of best quality supportive and palliative care for people nearing the end of their life, so that they have better living in the last year(s) of their life, and better dying.

The objectives of the 4CEHR in the clinical sphere were:

- using a shared electronic health record, to improve the health assessment of elderly people nearing the end of their life;
- to make RACF residents' ACPs and CAPs available in a 4CEHR to improve communication, collaboration and coordination with Primary Health Care Teams, including general practitioners and RACF staff, acute hospitals, allied health, after-hours services such as GP Assist, and other specialist providers including the Specialist Palliative Care Service, the Older Persons Mental Health Service and geriatricians.

To deal with the lack of specific requirements for the system, the project manager and vendor adopted an agile method to develop the system incrementally. In an attempt to develop the first prototype, existing ACD forms in use in Australia supplemented by those proposed by the clinical project team were used to identify themes and common elements. Iterative development was subsequently enabled based on the feedback of users.

Recognising user input was critical for iterative development in this uncertain and fluid development environment, but that users were time-poor and development time was constrained, a Thinking Out Loud technique was adopted to supplement a more conventional user requirement stage of systems development. We proposed that timely evaluative feedback could be obtained to support agile development using Thinking Out Loud, which is normally used when performing systems evaluation (Nielsen 1994)

## 3  METHOD

The research method adopted was that of a single Interpretivist qualitative case study. The research design consisted of two components: agile systems development and user feedback; Thinking Out Loud: Iterative Requirements Gathering through Usability Testing

Usability evaluations were deployed on all iterations of the prototype. An external researcher outside of the project team conducted these evaluations. This independence during the evaluation was deemed to be important; the researcher was unknown to the users and had limited detailed understanding of the system or process being evaluated. The method adopted for the usability testing was *Thinking Out Loud* (Kushniruk and Patel 2004). This method consists of two stages:

(1) Collecting Thinking Out Loud protocols in a systematic way, and

(2) Analysing these protocols to obtain a model of the cognitive processes that take place while a person tackles a problem (Ericsson and Simon 1993).

This is a simple method where the user is asked to use the system based on a scenario that has been provided. The user is given time to read and understand the scenario. The aim is for the user to verbalise, out loud, their thoughts and reflections while attempting to complete the task as described in the scenario. *Thinking Out Loud* process commences prior to the user opening the system through to completion of the task allocated.

Traditionally the information is captured using screen-capture software, audio and video recordings and is later analysed by the project team. The analysis is in-depth and resource intensive. In this project the process that was adopted was a refined (lite) version of Thinking Out Loud which better suited the lack of requirements modelling and the extremely tight project timelines. The *Thinking Out Loud Lite* involved:

- 2 researchers as observers;
- screen capture software;
- scenario.

The two researchers observed the user while the evaluation was being conducted. The user was reassured that this was not a test and that the observations were focused on how they used the system. One researcher was positioned next to the user so they could observe both the user and the system. This researcher also prompted the user to speak if they fell silent while exploring the system. The second researcher was positioned somewhere else in the room so as not to overwhelm the user during the evaluation. Notes were made by each of the researchers paying particular attention to visual and verbal cues from the user. The observation notes formed the primary source for data analysis.

CamStudio screen-capturing software was employed to capture screen movement (http://camstudio.org/). As the evaluations took place on site at four Registered Aged Care Facilities (RACFs) which involved different types of infrastructure and varied levels of technology, relying on the capture software as the primary data source was considered too risky. The screen-capture software was employed as a backup to the researchers' observational notes.

The scenario was developed by the project manager to reflect an activity that the users would encounter while admitting a patient and then later gathering information relating to them and their wishes. Users were provided with a scenario to read relating to a new patient being admitted into the RACF. The scenario provided details on the patient's health, his immediate wishes, next of kin, and if registered, their Enduring Guardian. The scenario provided the basis for the users to enter a new record into the system.

At the completion of the usability testing a 2-hour brainstorming and analysis session was conducted with the observers and the project manager. This approach is in line with Instant Data Analysis (IDA) developed by researchers in Denmark (Kjeldskov et al. 2004). The aim of this approach is to be able to provide fast identification of the most critical usability problems of a software system. The issues revealed in the analysis were categorised and ranked according to their severity.

Severity was categorised as:

- Usability catastrophe - imperative to fix this;
- Major problem - important to fix;
- Minor problem – Fixing should be given a low priority;
- Cosmetic problem - need not be fixed.

The issues are also categorised into theme areas:

- Guidance – availability to advise, orient, inform, instruct and guide the users through their interaction with the computer;
- Workload – interface elements that play a role in the reduction of the users' perceptual or cognitive load, and in the increase of the dialogue efficiency;
- Explicit Control – it concerns both the system processing of explicit user actions, and the control users have on the processing of their action by the system;
- Adaptability – capacity to behave contextually and according to the users' needs and preferences.
- Error Management - means availability to prevent or reduce errors and to recover from them when they occur;
- Consistency – it refers to the way interface design choices are maintained in similar contexts and are different when applied to different contexts;
- Significance of codes – relationship between a term and/or a sign and its reference;
- Compatibility – match between users' characteristics and task characteristics on the one hand, and the organisation of the output, input, and dialogue for a given application, on the other hand;

## 4　RESULTS

The outcome from the first evaluation of the prototype is provided in Table 2 as an example. Issues varied from: 'save' not clear and requires a prompt command; duplication of data across screens; not sure of what needs to be done first, slider needs to be darker; entry to patient record not clear; lack of logical arrangement; tab button works intermittently; all address fields require restructuring; spell check required.

|  | Usability Catastrophe | Major Problem | Minor Problem | Cosmetic problem |
|---|---|---|---|---|
| Guidance | 0 | 7 | 5 | 1 |
| Workload | 0 | 4 | 2 | 0 |
| Explicit Control | 0 | 1 | 0 | 0 |
| Adaptability | 0 | 1 | 0 | 0 |
| Error Management | 1 | 5 | 0 | 0 |
| Consistency | 0 | 2 | 0 | 0 |
| Significance of codes | 0 | 1 | 0 | 0 |
| Compatibility | 0 | 1 | 1 | 0 |

*Table 2. Number of issues ranked by severity and category*

In general the users were impressed with this first iteration of the 4C system. Expectations had been managed as users were repeatedly told that this was the first iteration of the system and it had been modelled on the existing paper forms. The main outcome of the first evaluation was that the project team needed to look at the logical gathering and flow of information along with auto-population of fields.

## 5   DISCUSSION

When users engage with information system professionals and begin exploring requirements, aspirations and different insights into what is needed emerge. In healthcare this has traditionally given rise to significant boundaries which have made information sharing difficult as "each group coalesces around divergent orientations towards healthcare delivery" (Randall and Munro 2010). The tendency for professionals to undervalue the other has also been noted (Bernstein et al. 2011). In order to represent a new model of integrated care delivery which transcended professional and organisational boundaries, the delivery of a prototype early was crucial in enabling users to explicitly see what was until then a concept. *Thinking Out Loud Lite'* was a useful tool with which to test an early prototype with users that did not require a huge investment both in terms of their time and the time required for development. Major operational issues were identified during the design phases with further design ideas generated and incorporated.

Thinking Out Loud assisted in testing the 4C system against objectives in the personal and clinical spheres. The extent of illness and limited capacity of the residents of the RACF precluded their involvement in usability testing. By recruiting testers from across the multidisciplinary team we were able to test elements of the system designed to meet the objectives in the personal sphere. For example, nursing staff modelled registering a resident on the system when they were admitted to the RACF, with the scenario describing certain patient preferences for care, inclusion of person responsible or Enduring Guardian, or other resident wishes. In the clinical sphere users from the primary health care team tested the system within the role they performed using different scenarios – for instance, one scenario required an afterhours GP to logon to a patient record, and another required finalising care plans. Within the population sphere the 4C system was developed using national infrastructure and standards where available, but these were technical specifications unsuited to the Think Out Loud approach and were tested in an external environment. Advice was provided to project sponsors to inform further refinement of components of the national PCEHR infrastructure. Patient-centred objectives within the population sphere were longer term goals of the system for future evaluation.

### 5.1   Implications for eHealth development

Iteration is inevitable because designers never get the solution right the first time (Gould and Lewis 1995, Preece et al. 1994). The impost on user participants can be large, and objections to having to find time for testing are common (Andrzeevski 2010; Franklin 2008). This is particularly challenging in healthcare organisations where practitioners are typically time-poor. *Thinking Out Loud Lite* provided an efficient way of focusing user time, with preparatory scenario building and analysis of results completed by project staff. The richness of the available video data is seductive and can consume an inordinate amount of vendor time if the principles of the process and reportable criteria are not agreed beforehand. Having project team members with skills in both domains to foster the collective intelligence of the collaborative user group, understand the contextual elements of the users' needs, and be able to communicate this to developers is important (Chi et al. 2010; Gregorio 2012; Preece et al. 2011).

Agile development would seem suitable to building information systems which align with new models of healthcare delivery, where the end product cannot be known until at least part of the solution is built (Preece et al. 2011). It involves iterative development throughout, where requirements are explored and tested. This requires some contract flexibility which is not well supported by government organisations where standard procurement practices seek greater control over project schedule, cost and deliverables (Andrzeevski 2010; Hoda et al. 2009). Typically, procurement practices favour the waterfall method where requirements are set and fixed prices attained prior to development. (Andrzeevski 2010; Franklin 2008)  To support reforms in public healthcare, federal and state government organisations need to find ways to adapt, using more flexible development contracts while maintaining due diligence.

### 5.2 Implications for research

This paper provides insight into a research project that was from the outset constrained by the difficulty in gathering user requirements along with restrictive time constraints for the delivery of the system. Often research projects are faced with these and other constraints that impede the progress of the project and the ensuing research. From a methodological viewpoint this paper has presented an insightful research design to overcome what in the past has been thought to be insurmountable constraints. Agile systems development is not new to systems development methodologies however aligned with *Thinking Out Loud Lite* the two create a powerful design for rapid development of a system that is guided by users as an integrated component of the iterative cycle. The three spheres of participation provided a useful framework to articulate objectives for meaningful use at the proposal stage, which might subsequently be helpful in evaluation.

Evaluation would ideally be conducted by suitably skilled individuals. The socio-technical domain is complex, with health and medico-legal issues further complicating the subject area. Social scientists have a role in understanding users, their motivation, and how this changes over time, but may have limited understanding of the technical and health domains. Challenges include misaligned methodological incentives, evaluation expectations, double standards, and relevance compared to industry (Bernstein et al. 2011).

### 5.3 Limitations and future work

While development was able to progress, iterations beyond what was possible in the project time period would have delivered a better quality system. The methodology described may be useful when challenges arise, but does not negate the need for requirements gathering or replace realistic development timeframes.

This project was a small project conducted in a rural community that provided easy access to end-users. The method proposed has not been tested on a larger project with a more diverse user group.

Ultimately the 4C system was decommissioned in 2015. Integration with existing user systems was never achieved due to major changes in stakeholder priorities – the proposed system within the hospital was deferred, and GP software vendors were contracted to deliver other components of the PCEHR as a priority. A system designed to support cross-boundary care will only add value if the computer infrastructure supports information flow between spheres seamlessly.

## 6 CONCLUSION

Traditional project management approaches used in Information Technology implementation judge project success in terms of measurable outcomes - meeting objectives, on time, within budget, and according to the sponsors' satisfaction. To derive business value from eHealth investment, systems should not just bridge existing silos but enable better ways of working in the context of the patient journey. This can be discovered during collaborative planning and development when existing knowledge and knowledge creation cultivate innovation. Flexible, collaborative tools to aid development are required. We found early delivery of a prototype and iterative requirements gathering and usability testing with *Thinking Out Loud Lite* supported this style in a timely manner enabling delivery of the project on schedule and within budget.

When developing systems for healthcare, understanding users is not just good practice it is critical to ensuring systems are safe and support health professionals, patients and their families to experience high quality and sustainable care. Healthcare settings often exhibit circumstances where key requirements for an e-Health system are not, (and/or should not) be defined at the outset. Given the importance, potential, and huge investment that Australia has made in the PCEHR it is disappointing to note the limited engagement with end-users and the failure to deploy well understood frameworks, tools and techniques that have proven value within this domain (Almond et al. 2013). This paper has highlighted a straight-forward technique that actively engages end-users in both iterative requirements gathering and usability testing to deliver an e-Health solution. Nevertheless, integrating systems seamlessly is less straight-forward and is critical if healthcare systems are to transform healthcare in the context of the patient journey.